**Critical Thinking with Generative AI: A Constraint-First Design Pilot of a Thinking-Partner Intervention**

Fatima Tuz Zahra[1], Jiangen He[2], David M. Bowers[3], Wei Wang[4]

[1]Department of Educational Leadership and Policy Studies, University of Tennessee, Knoxville, Bailey Education Complex, 1122 Volunteer Blvd, Knoxville, TN 37996, USA.

[2]School of Information Sciences, University of Tennessee, Knoxville, 1345 Circle Park Drive, Knoxville, TN 37996, USA.

[3]Department of Teaching and Learning, The Ohio State University, Columbus, OH 43210, USA

[4]University of Tennessee, Knoxville, Knoxville, TN 37996, USA

## Abstract

Generative AI (GenAI) tools entered higher education classrooms faster than the field was able to study their effects on learning. One concern is that GenAI may displace the critical thinking and AI literacy that students will need after graduation. This paper reports a Design-Based Research pilot of a GenAI-assisted critical thinking framework, in which ChatGPT was used as a thinking partner in an undergraduate research methods and statistics course during Spring 2025 (N = 14). The mixed-methods design combined pre- and post-intervention measures of statistical learning (AASCDM), AI literacy (MAILS), and critical thinking (WGCTA) with instructor field notes, student artifacts, and student-AI interaction logs. Pre-post tests showed gains on every AASCDM dimension and on eight of nine MAILS dimensions, while WGCTA percentiles did not change. Qualitative analysis identified four themes: the ways students positioned the LLM (as answer generator, validator, or co-thinker); the depth of student engagement (procedural vs. conceptual); occasional humanizing of the tool; and the role of curriculum design in shaping each of the prior three. Read together, the findings indicate that one semester of GenAI-assisted instruction can move domain learning and self-reported AI literacy but does not move standardized critical thinking, and that the modal student-LLM relationship is one of validation instead of dialogue. We end with design principles for the next iteration of the framework and implications for research on adaptive and personalized learning.

## 1 Introduction

The swift spread of artificial intelligence (AI) technologies, and of generative AI models such as ChatGPT in particular, has prompted broad discussion within education. As these technologies continue to develop, they show potential across educational applications, and researchers, educators, and policymakers have begun to examine the opportunities, challenges, and ethical questions involved in their use for teaching, learning, and administration (Bonner et al., 2023; Duha, 2023; Holmes et al., 2019; Kim et al., 2022; Lim et al., 2025; Su & Yang, 2023; Tack & Piech, 2022; Zawacki-Richter et al., 2019). The opportunity and the risk are both real, and the success or failure of any given implementation rests on a complicated set of pedagogical and contextual factors. As with prior technologies such as the calculator, AI-driven interventions can be designed in ways that either build or substitute for skills like critical thinking and AI literacy.

Alongside the spread of these technologies, employers and scholars have noted a gap between students' critical thinking (CT) abilities (NACE, 2023) and the demands of the modern workplace (World Economic Forum, 2025). Researchers have raised concerns about the inadequate development of CT skills among college students (Desai et al., 2016; Flores et al., 2010; Roohr & Burkander, 2020; Zembylas, 2022) and have argued for educational models that prioritize intellectual engagement and problem-solving.

Addressing this challenge in situ with AI is still an emerging area in literature (Schiff, 2021), and it raises questions about the pedagogical mechanisms that shape students' ability to think critically and independently in technology-mediated environments. Generative AI (GenAI) is a candidate tool for this work, but the risk of AI weakening CT instead of strengthening it is real, given that overreliance on AI tools can substitute for analysis and promote intellectual complacency (Guo & Lee, 2023; Konak & Clarke, 2023). As with the calculator example, learners need AI literacy in order to use AI in ways that build on their own thinking instead of replacing it.

This paper reports findings that begin to identify the conditions under which an AI-augmented pedagogical framework supports student learning. The findings come from a Design-Based Research (DBR) study (Brown, 1992; Collins, 1992; Collins et al., 2004) that used ChatGPT as a personalized, adaptive thinking partner in support of students' critical thinking and AI literacy. We piloted the initial version of the framework in an undergraduate research methods and statistics course during Spring 2025. The paper is organized around three research questions:

RQ1. To what extent did students show pre-post gains in statistical learning, AI literacy, and critical thinking after participating in the GenAI-assisted intervention?

RQ2. How did students position and engage with the GenAI tool over the course of the semester?

RQ3. What can we learn from students' engagement with the GenAI tool related to learning outcomes and to ongoing design refinements?

## 2 **Literature Review**

This review draws on three areas of work: critical thinking development in higher education, AI literacy frameworks, and the role of conceptual versus procedural knowledge in technology-mediated learning. Together, these literatures govern the design and analytic choices in the pilot reported here.

## 2.1 Critical Thinking in Higher Education

Critical thinking is among the most widely endorsed educational outcomes in higher education, even as its definition and assessment remain contested. The American Philosophical Association's Delphi Project (Facione, 1990) produced a consensus definition that continues to inform contemporary research: critical thinking is purposeful, self-regulating judgment yielding interpretation, analysis, evaluation, and inference, accompanied by an explanation of the evidential, conceptual, methodological, or contextual considerations on which that judgment is based. The definition treats critical thinking as both a learnable set of cognitive skills and a disposition toward thoughtful inquiry. Anderson and Krathwohl's (2001) revised Bloom's taxonomy provides a complementary anchoring, positioning critical thinking within the higher levels of the cognitive hierarchy where students analyze, evaluate, and create instead of remembering or applying (Krathwohl, 2002).

Despite this shared definition, disagreement remains. Reviews note continuing debate about how to define and measure CT (Niu et al., 2013), with philosophers emphasizing logical reasoning, psychologists focusing on cognitive processes, and educators prioritizing pedagogical applications. Ennis (1987, 2011) drew a complementary distinction between critical thinking as cognitive skills and critical thinking as dispositions, with research showing only moderate correlation between the two facets (Facione et al., 2000).

Assessment is similarly contested. The California Critical Thinking Skills Test (CCTST) is among the most widely used CT instruments and has shown sensitivity to instructional gains (Facione, 1990). Critics, however, have raised concerns about circularity in its validation logic and about the formal-logic items that may not capture the messier reasoning required in authentic contexts (Bailin, 2002). The Watson-Glaser Critical Thinking Appraisal (WGCTA), used in the present study, focuses on inference, recognition of assumptions, deduction, interpretation, and evaluation of arguments (Bernard et al., 2008), and is widely used in higher education and workplace settings (Watson & Glaser, 2010).

The pedagogical challenge compounds these measurement issues. Students often arrive in higher education either presuming that they will not be required to think and question, or actively avoiding doing so (Burns et al., 2018; O'Leary & Scully, 2018). Instructors are therefore in the position of both undoing years of conditioning toward passive learning and teaching skills whose definitions are still under debate. Meta-analytic evidence indicates that direct, explicit CT instruction with mentored argumentation and justification tasks produces measurable gains in critical thinking, but that transfer of those gains across domains requires deliberate design of the instructional sequence (Abrami et al., 2008, 2015; Halpern, 1998).

## 2.2 AI Literacy: Emerging Frameworks

AI literacy has emerged as a distinct educational priority that responds to both technological change and conceptual reorganization. Unlike traditional digital literacy, AI literacy refers to the knowledge, skills, and dispositions needed to understand, evaluate, and use AI technologies responsibly and effectively (Long & Magerko, 2020).

The EDUCAUSE framework organizes AI literacy into four areas: technical understanding of how AI works, evaluation of AI tool applications and outputs, application and management of AI tools, and ethical strategies for guarding against bias and misuse (Kassorla et al., 2024). The Stanford Teaching Commons framework emphasizes the rhetorical and communicative dimensions of AI literacy, including how to use language strategically with AI tools (Stanford Teaching Commons, 2024). Digital Promise's AI Literacy Framework adopts a process-oriented approach with three modes (understand, evaluate, use) and stresses that understanding and evaluating AI are central to making informed decisions about whether and how to use AI in learning environments (Digital Promise, 2025).

These frameworks share an assumption: that the evaluation of AI is itself a critical thinking task. The assumption is reasonable but recursive. If our conceptualization and measurement of critical thinking remain unsettled, then the part of AI literacy that depends on critical evaluation is built on contested ground. We treat this as a reason to study AI literacy and critical thinking together, not as separate outcomes.

The Meta AI Literacy Scale (MAILS; Carolus et al., 2023), used in the present study, measures AI literacy across nine dimensions (Apply AI, Understand AI, Detect AI, AI Ethics, Create AI, AI Problem Solving, AI Learning, AI Persuasion Literacy, and AI Emotion Regulation). MAILS has been validated for self-report use with adult learners and provides finer granularity than earlier unidimensional scales.

### 2.3 GenAI and Critical Thinking: Empirical Evidence

The empirical literature on generative AI and critical thinking complicates straightforward claims about AI as an educational accelerant. AI systems are contemporary tools for substituting cognition that require minimal participation of users, and can still provide comprehensive results (Jose et al., 2025). Lee et al. (2025) surveyed 319 knowledge workers about their GenAI use and found that higher confidence in GenAI was associated with less critical thinking, while higher self-confidence was associated with more critical thinking. The same study reported that GenAI shifts the nature of critical thinking from generative work toward verification of information, integration of responses, and oversight of tasks (Lee et al., 2025). Instead of removing critical thinking from the workflow, GenAI appears to redistribute it. Constant access to AI-assistive technologies can improve humans' efficiency and work performance, yet it is important to consider the cost of cognitively offloading tasks on a consistent basis (Gerlich, 2025; Lee et al., 2025). Recent work in scientific research has documented an analogous concern: AI-mediated analysis can produce an illusion of understanding that masks gaps in genuine comprehension (Messeri & Crockett, 2024), a caution that applies with equal force to student learning.

Educational research has begun to document analogous effects in formal learning. Gonsalves (2024) found that AI can both enhance and challenge critical thinking across cognitive, affective, and metacognitive domains. AI-powered tools can be used to augment critical thinking skills, and it is determinant on the pedagogy and teaching methods surrounding its use (Hassen, 2025). AI can support creativity and problem-solving by creating dynamic, interactive systems for students, as well as aid in reading comprehension (Hassen, 2025). Nevertheless, improper regulation poses a risk to students' cognitive development and can foster intellectual apathy if safeguards are not implemented. Overreliance is a particular concern: when users accept AI-generated recommendations without questioning them, performance errors increase in decision-making contexts (Zhai & Wibowo, 2024). The pattern is more than a training problem. It may reflect features of human-AI interaction that resist short-term pedagogical correction.

Users develop inappropriate trust in AI systems precisely because those systems are often correct and helpful, which makes occasional failures harder to catch. This trust can lead to a habitual dependence, to delegate making critical choices and evaluations to AI systems without human validation, and discourage users from any in-depth engagement (Ejaz et al., 2025;

Gerlich, 2025). Extensive use of AI for decision making can contribute to a decline in divergent thinking if individuals continue to rely on algorithmically produced recommendations, which will eventually erode people's ability to explore unconventional solutions (Sireen et al., 2026). Prolonged reliance on AI to solve complex problems can lead to users minimizing the use of their own critical thinking, and lowers their threshold to handle mental fatigue (Tian & Zhang, 2025).

## 2.4 Constructivist Theory and Technology-Mediated Learning

Constructivist learning theory provides a useful frame for understanding knowledge construction, and its application to AI-assisted learning environments raises tensions of its own. McLeod (2025) describes constructivism as both a learning theory and a philosophy of education in which learners actively build knowledge through experiences and interactions. The foundational principles emphasize active knowledge construction over passive transmission (Mattar, 2019). AI tools complicate the active/passive distinction by providing information processing that, from the learner's view, can resemble understanding. When AI systems generate explanations, analyses, or solutions, learners may experience cognitive engagement without producing knowledge construction of their own.

Research on technology-mediated constructivist pedagogy has reported positive correlations between such pedagogy and student engagement (Allen, 2022). Behavioral engagement metrics (time on task, interaction frequency, completion rates) do not necessarily map onto meaningful learning, however. This is one reason the present study pairs validated learning measures with interaction analysis and qualitative observation, not on engagement indicators alone.

Vygotsky's zone of proximal development (ZPD) is particularly relevant in AI-assisted contexts. Traditional ZPD assumes that learning occurs through interaction with more knowledgeable others who provide appropriate scaffolding. AI systems can produce sophisticated scaffolds, but they lack a model of the individual learner's needs, misconceptions, or developmental trajectory. They offer assistance without comprehension, a caution grounded in critiques of large language models as systems that generate fluent text without semantic understanding (Bender et al., 2021). This has implications for how educators must design around them.

### 2.5 Conceptual and Procedural Knowledge in AI-Assisted Learning

The distinction between conceptual and procedural knowledge offers a useful frame for examining AI's educational role. Hiebert and Lefevre (1986) described procedural knowledge as action sequences and conceptual knowledge as understanding of principles and relationships. The pedagogical principle that conceptual knowledge should anchor procedural fluency (Grouws & Cebulla, 2000; Hurrell, 2021) becomes more complicated when AI tools can supply both at once.

Lockhart's (2002) musician metaphor cautions against reducing understanding to procedural execution, comparing it with turning music into notation exercises that ignore sound itself. The same caution applies to AI-assisted learning: tools that produce well-articulated conceptual explanations can lead students to mistake AI-generated articulation for their own understanding. The risk is not procedural reductionism alone but a new pattern in which students articulate sophisticated explanations they have not internalized. Metacognitive pedagogy, which makes the act of thinking itself an object of instruction, offers one route out of this pattern (Flavell, 1979; Tanner, 2012). Sidra and Mason (2024) extend this logic to AI literacy specifically, arguing that metacognitive thinking should be treated as a defining component of AI literacy in the workforce, not as an adjacent skill.

### 2.6 Design-Based Research as Methodology

Design-based research (DBR) is a methodology suited to investigating complex educational interventions, including those involving emerging technologies. Scott et al. (2020) describe DBR as the study of learning ecologies that are grounded in theories of learning, produce measurable changes in student learning, generate design principles to guide instructional tools, and are enacted through extended, iterative teaching experiments. DBR fits the present problem because the questions are both empirical (does the intervention work?) and design-oriented (what should the next iteration look like?). It supports the iterative refinement that emerging technologies demand and accommodates the integration of quantitative and qualitative evidence.

### 2.7 Synthesis and Gap

Three gaps in the literature motivate the present study. First, existing research often treats critical thinking and AI literacy as parallel constructs without examining how they interact in authentic learning contexts. Second, most empirical studies focus on short-term behavioral indicators without pairing them with validated learning measures and direct observation of AI interaction. Third, the literature has relatively few iterative, design-oriented studies that document outcomes alongside the design moves that produce them.

The pilot reported here begins to address these gaps. We pair validated assessments of statistical learning, AI literacy, and critical thinking with analysis of student-AI interactions and qualitative observation, situated within a DBR cycle that yields design principles for subsequent iterations.

## 3 Methodology

### 3.1 Design-Based Research

Design-based research has proven useful for closing the gap between research and classroom application in formal education (Anderson & Shattuck, 2012). Beyond measuring framework efficacy, DBR yields empirical insights for optimizing the design and integration of GenAI in education (Lim et al., 2025). DBR interventions are evaluated using diverse methodological approaches and multiple assessment indices, making mixed methods particularly suitable for these investigations (Bell et al., 2013). Grounded in the foundational work of Brown (1992) and Collins (1992), Reeves (2000, 2006) outlines key principles for DBR: it should (1) address complex, real-world problems in partnership with practitioners, (2) integrate established and theoretical design principles with technological affordances to develop viable solutions, and (3) employ rigorous, iterative inquiry to refine innovative learning environments and generate new design principles. Guided by this framework, the present study adopts a DBR approach with mixed methods to examine the impact of GenAI as a thinking partner on students' critical thinking, AI literacy, and statistical learning.

### 3.2 Context and Implementation

#### 3.2.1 Study Setting

The pilot was implemented in Spring 2025 in an undergraduate applied statistics course for decision making at an R1 public university. The course met for one semester and enrolled 14 students who completed both pre- and post-intervention measures. The institution serves a regionally and socioeconomically diverse student body and offers reliable institutional access to GenAI tools, removing access as a confound in our analysis. The Institutional Review Board approved the study (IRB-23-07810-XP), and all participants provided informed consent.

### 3.2.2 The Thinking-Partner Implementation: A Constraint-First Design

The three-step interaction protocol used here — students first think on their own, then engage with the AI, then report what shifted in their thinking — was introduced by the first author in a prior conference presentation (Author, 2025). The pilot reported in this paper is the first empirical implementation of the protocol in a semester-long course. The intervention positioned ChatGPT as a thinking partner in place of an authoritative source of answers (Tang et al., 2024; Tang & Putra, 2026) and assigned to this role in recurring weekly course activities structured around guided, iterative refinement (Lee et al., 2024; Lin et al., 2024). We use "thinking partner" throughout this paper to name the intended cognitive relationship between student and LLM; the term is synonymous with "dialogic partner" as used in the system prompt shown in Appendix A. The resulting design rested on three choices. First, the LLM was framed as a thinking partner instead of an answer source: course assignments asked students to use the LLM to test, refine, and stress-test their own arguments, not to produce them. Second, instructors gave the model an initial prompt and context window intended to give the model a warm, Socratic register, and students received guidance on how to refine their prompts as their questions became more specific. Third, instructors repeatedly returned to a frame summarized in class as “smart humans, simple machines,” to remind students that the LLM operates statistically and lacks the semantic grounding that human thinkers bring to a problem (Bender et al., 2021), and that responsibility for reasoning therefore remains with the learner.

The implementation ran through two parallel channels: homework assignments and weekly practicum sessions. The course was structured around two class meetings each week, one focused on lecture and discussion, and one focused on applied practice. The thinking partner was reinforced in the practice-focused meeting, where students worked through problem sets alongside the LLM and where documentation of student-AI interaction was made a routine part of the classroom process. Homework assignments were problem-solving focused and required students to make methodological decisions with the thinking partner. Every assignment carried a rubric that included a dedicated AI partnership category worth approximately five points on average, which credited the quality of the student's use of the thinking partner, not the quality of the LLM's output. For both homework and classroom, students submitted their written work along with the corresponding chat logs on Canvas, which produced the interaction record used in Section 3.4. Beginning in Week 3, the weekly included a class activity in which Assignments 2-8 provided seven opportunities for the homework-based engagement with the thinking partner, of which five were graded assignments.

Two design revisions emerged over the semester in response to the qualitative analysis described in Section 3.5. The first was pedagogical: instead of expecting students to arrive at the co-thinker orientation on their own, the practicum sessions were adjusted to rehearse the same problem-solving process that would appear in the incoming assignment, so that students could develop a working rhythm for engaging with the thinking partner before facing a graded task. The second was epistemic: instructors began reinforcing more explicitly and more frequently that the student's own thinking and decision-making mattered more than what the LLM produced, and that the LLM's output was one input to the student's reasoning instead of the destination of it.

### 3.3 Quantitative Methods

Central to survey research is operationalization, the process of translating theoretical constructs into measurable variables (Nardi, 2018). To assess the effect of the intervention on students' critical thinking, AI literacy, and statistical learning, we used a pre-post design with three validated instruments and a demographic questionnaire.

The Watson-Glaser Critical Thinking Appraisal (WGCTA; Watson & Glaser, 2010) was used to assess core reasoning abilities, generating subscale percentiles for Drawing Conclusions, Evaluating Arguments, and Recognizing Assumptions (Bernard et al., 2008). The Meta AI Literacy Scale (MAILS; Carolus et al., 2023) is a 34-item questionnaire measuring nine AI literacy competencies (Apply AI, Understand AI, Detect AI, AI Ethics, Create AI, AI Problem Solving, AI Learning, AI Persuasion Literacy, and AI Emotion Regulation) on an 11-point Likert scale. The Applied Statistical Concepts for Decision Makers assessment (AASCDM) is a 10-item self-report instrument developed by the first author for this study to assess perceived mastery of the statistical methods covered in the course, with two items on GenAI tool use; items were rated on a 3-point confidence scale. Internal consistency was acceptable at pretest (Cronbach's $\alpha = .73$). At posttest, a ceiling effect (item means ranging from 2.36 to 3.00) reduced item variance and lowered $\alpha$ to .51, a limitation we return to in Section 5. Together these instruments yield data on intervention effects across cognitive and technical domains.

To address RQ1, we used paired-samples t-tests on AASCDM and MAILS dimensions, and Wilcoxon signed-rank tests on WGCTA subscale percentiles given the small sample size and non-normal distribution. To address RQ3, we used descriptive analyses to examine relationships between students' AI interaction patterns (query frequency, complexity, and sentiment, described in Section 3.4) and learning outcomes. Given N = 14, we report exact test statistics and treat the statistical results as descriptive of this pilot in place of confirmatory evidence.

This quantitative approach is integrated with the qualitative methods described in Section 3.5 within a broader mixed-methods design. It is intended both to indicate whether the intervention is working and to surface where it is and is not working in ways that inform the next DBR iteration.

### 3.4 Human Behavior Analysis

To complement the validated outcome measures, we analyzed logs of students' interactions with the AI thinking partner across the semester. The corpus comprised 442 student-authored prompts from seven assignments, averaging 197.4 words (1,252 characters) per message, with 36 to 92 prompts per assignment (M = 63.1). Logs were de-identified and stored on secure institutional servers in accordance with the IRB-approved data management plan. Course assignments comprised 194 discrete questions across the seven assignments, with 74 questions (38%) tagged as reflection prompts intended to elicit higher-order thinking (see Appendix B). Three complementary analyses addressed what students discussed, what they asked the AI to do, and how their requests unfolded across turns.

*Topic analysis.* We embedded each prompt with OpenAI's text-embedding-3-large model and clustered the resulting vectors with BERTopic (Grootendorst, 2022), using HDBSCAN with parameters tuned for short text (minimum cluster size = 15, minimum samples = 5, cluster selection epsilon = 0.3). GPT-4.1 served as the representation model, generating interpretable labels from each cluster's most representative documents and keywords. UMAP (McInnes et al., 2018) projected the embeddings into two dimensions for visual inspection.

*Prompt categorization.* Two researchers developed a coding schema iteratively from a random sample of 50 prompts. A second random sample of 50 prompts was then coded independently by two coders using that schema, and a third coder adjudicated every disagreement, producing a consensus-coded gold standard. We classified the full corpus with GPT-5, prompted with category definitions, worked examples, and decision criteria, and constrained to return a JSON object containing a category label and a confidence score from 0.0 to 1.0 (system prompt in Appendix A). The full category codebook with definitions and anonymized student examples is provided in Appendix C. Classifier agreement with the gold standard is reported in Section 4.2.

*Sequential pattern mining.* We represented each conversation as an ordered sequence of category labels and computed n-gram frequencies for bigrams, trigrams, and fourgrams by sliding a window of size $n$ across each sequence. We also estimated a first-order Markov chain in which cell $(i, j)$ gives the probability of moving from category $i$ to category $j$, calculated as the frequency of $i \rightarrow j$ transitions divided by the total frequency of category $i$. Together these analyses characterize not only what students asked but how their requests were ordered and how one type of request led to another.

### 3.5 Qualitative Methods

DBR, given its goal of productively merging the abstractions of research with the pragmatics of the classroom, draws on mixed methods to offer a multidimensional view of objects of analysis. Here it is useful to think of the quantitative methods as summative and the qualitative methods as formative. Whereas the quantitative methods provide insight into the overall efficacy of our pedagogical implementation, the qualitative methods reported here characterize and guide what aspects of the implementation might need refinement while we remain in the midst of ongoing refinements. DBR refinements are iterative; here we report primarily on what our qualitative analysis revealed during this pilot implementation and on the refinements that analysis prompted. A more substantial retrospective qualitative analysis is reported separately.

Across the semester, we engaged in continuous iterations of the DIER design cycle, ultimately producing two major revisions. These revisions arose as we (1) applied critical thematic analysis (Lawless & Chen, 2019) to assignments, student homework assignments, and classroom observations through the conceptual frame of critical landscapes of education (Penteado & Skovsmose, 2022; Skovsmose, 2023, 2024) and (2) followed up with finer-grained analysis of the themes that surfaced. For clarity, we report here on the second analysis, an evaluation of content and implementation through the lens of conceptual and procedural knowledge, and we share how those findings guided ongoing design revisions. The goal of the qualitative component is to make visible a lived story of design: what revisions did we make, when, and what prompted them. These details enable a more robust analysis of our research findings and create space for nuanced insights into implications for future work at the intersection of AI and education.

## 4 Findings and Discussion

### 4.1 Quantitative Findings

#### 4.1.1 Sample Demographics

Fourteen students participated in the study (N = 14). All participants were undergraduate students. The majority (approximately 65%) were enrolled in the education-focused college that hosted the course, with smaller numbers from other colleges including arts and sciences and communication and information. Class standing ranged from freshman to junior level, with most participants in their first or second year (Table 1).

Most participants identified as women (approximately 79%), with the remainder identifying as men. All respondents were between 18 and 20 years old. About one third of the sample (36%) self-identified as neurodiverse, and 21% reported having a disability. None of the participants were first-generation college students. Most students (86%) were employed either part time or full time, and all respondents reported reliable access to a laptop or desktop and a stable internet connection. Self-rated digital literacy was high, with most participants describing themselves as advanced users of digital tools. Reported household incomes ranged from approximately $40,000 to over $200,000, with the median bracket falling between $100,000 and $149,999. None of the respondents reported having dependents. The sample was therefore technologically well-equipped, academically early-stage, with relatively high digital competence and upper-middle-income socioeconomic backgrounds.

Table 1

Participant demographic characteristics

| **Characteristic** | **Distribution** |
| --- | --- |
| Enrollment status | Undergraduate: 14 (100.0%) |
| Age | 18-20 years: 14 (100.0%) |
| College | Education, Health, and Human Sciences: approximately 9 (64.3%); other colleges: 5 (35.7%) |
| Class standing | Freshman: 10 (71.4%); sophomore: 1 (7.1%); junior: 3 (21.4%) |
| Gender | Women: 11 (78.6%); men: 3 (21.4%) |

| Characteristic | Distribution |
|---|---|
| Neurodivergence | Yes: 5 (35.7%); no: 9 (64.3%) |
| Disability | Yes: 3 (21.4%); no: 11 (78.6%) |
| First-generation status | Yes: 0 (0.0%); no: 14 (100.0%) |
| Employment | Part time: 10 (71.4%); full time: 1 (7.1%); not employed: 3 (21.4%) |
| Digital literacy | Advanced: 8 (57.1%); intermediate: 6 (42.9%) |
| Technology access | Computer access: 14 (100.0%); reliable internet: 14 (100.0%) |
| Dependents | None: 14 (100.0%) |
| Household income | Range: approximately $40,000 to more than $200,000; median category: $100,000-$149,999 |

4.1.2 AASCDM (Statistical Learning) Pre-Post Comparison

Paired-samples *t*-tests were conducted on students' self-assessment across ten learning dimensions (Table 2). Significant improvements were observed across all ten dimensions. Confidence in interpreting frequency distributions increased from $M = 2.21$ to $M = 3.00$, $t = -4.20$, $p < .01$, indicating an enhanced ability to comprehend data visualizations and distributions. Understanding of statistical variability also increased significantly ($M = 2.21$ to $M = 3.00$, $t = -4.20$, $p < .01$), and confidence in calculating and interpreting measures of central tendency improved notably ($M = 2.29$ to $M = 2.93$, $t = -3.80$, $p < .01$).

The largest gains were observed in selecting appropriate statistical tests ($M = 1.57$ to $M = 2.79$, $t = -6.50$, $p < .001$), using statistical software ($M = 1.07$ to $M = 2.36$, $t = -6.62$, $p < .001$), and explaining probability distributions ($M = 1.64$ to $M = 2.93$, $t = -7.87$, $p < .001$), suggesting that practical, software-oriented training significantly improved students' operational proficiency and conceptual clarity. Comfort in interpreting hypothesis test results also improved ($M = 1.86$ to *M* = 2.50, $t = -3.23$, $p < .01$). Students reported substantial growth in use of AI tools (*M* = 1.36 to *M* = 2.86, $t = -10.82$, $p < .001$) and perceived helpfulness of AI tools (*M* = 2.21 to *M* = 2.86, $t = -3.80$, $p < .01$).

Table 2

Pre-Post AASCDM paired-samples results

| Measure | Pre M (SD) | Post M (SD) | Change | *t*(13) | *p* | Holm *p* | d_z |
|---|---|---|---|---|---|---|---|
| Confidence with statistical terms | 2.21 (0.70) | 2.93 (0.27) | 0.71 | 3.68 | = .003 | = .009 | 0.98 |
| Interpret frequency distributions | 2.21 (0.70) | 3.00 (0.00) | 0.79 | 4.20 | = .001 | = .006 | 1.12 |
| Central tendency | 2.29 (0.61) | 2.93 (0.27) | 0.64 | 3.80 | = .002 | = .009 | 1.02 |
| Statistical variability | 2.21 (0.70) | 3.00 (0.00) | 0.79 | 4.20 | = .001 | = .006 | 1.12 |
| Select statistical tests | 1.57 (0.51) | 2.79 (0.43) | 1.21 | 6.50 | < .001 | < .001 | 1.74 |
| Use statistical software | 1.07 (0.27) | 2.36 (0.63) | 1.29 | 6.62 | < .001 | < .001 | 1.77 |
| Explain probability distributions | 1.64 (0.50) | 2.93 (0.27) | 1.29 | 7.87 | < .001 | < .001 | 2.10 |
| Interpret hypothesis tests | 1.86 (0.77) | 2.50 (0.52) | 0.64 | 3.23 | = .007 | = .009 | 0.86 |
| Use AI tools | 1.36 (0.50) | 2.86 (0.36) | 1.50 | 10.82 | < .001 | < .001 | 2.89 |
| Perceived AI helpfulness | 2.21 (0.58) | 2.86 (0.36) | 0.64 | 3.80 | = .002 | = .009 | 1.02 |

*Note.* Higher scores indicate greater confidence or endorsement on a 1-3 scale. Change was calculated as Post - Pre; positive t values therefore indicate higher posttest scores. d_z is the paired-samples standardized mean change.

Figure 1

AASCDM self-assessment scores at pretest and posttest

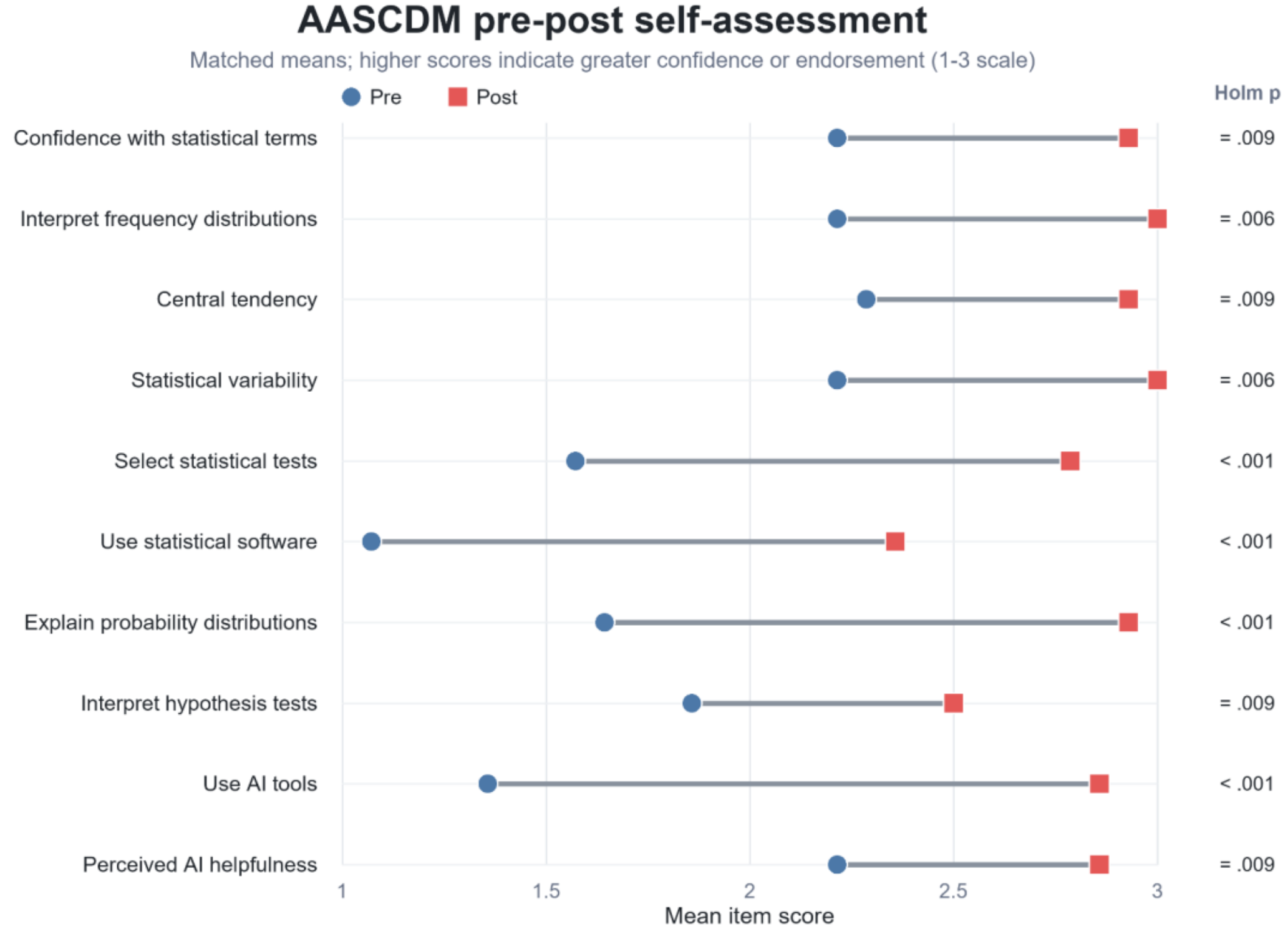

*Note.* Points show matched-sample means. Holm-adjusted p values are shown at right.

4.1.3 MAILS (AI Literacy) Pre-Post Comparison

Paired-samples *t*-tests revealed significant gains on eight of the nine MAILS dimensions (Table 3). Apply AI improved from $M = 36.64$ to $M = 53.00$, $t(13) = 5.82$, $p < .001$, and Understand AI improved from $M = 33.93$ to $M = 51.64$, $t(13) = 6.02$, $p < .001$. Detect AI showed significant improvement ($M = 17.57$ to $M = 24.71$, $t(13) = 3.49$, $p < .001$), as did AI Ethics ($M = 20.29$ to $M = 25.93$, $t(13) = 3.33$, $p < .01$). Notable gains were also observed in Create AI ($M = 6.07$ to $M = 20.43$, $t(13) = 4.64$, $p < .001$), AI Problem Solving ($M = 15.14$ to $M = 25.00$, $t(13) =$ 4.61, $p < .001$), and AI Learning ($M = 13.21$ to $M = 22.64$, $t(13) = 6.17$, $p < .001$). AI Emotion Regulation also improved ($M = 23.64$ to $M = 26.50$, $t(13) = 2.46$, $p < .05$).

The single dimension, which did not change significantly was AI Persuasion Literacy ($M = 25.43$ to $M = 25.86$, $t(13) = 0.32$, $p > .05$), suggesting that this dimension may require targeted educational strategies for cultivating critical evaluation of, and resilience toward, AI-driven persuasive influences.

Table 3

MAILS Pre-Post Comparison: paired-samples results

| **Dimension (items)** | **Pre M (SD)** | **Post M (SD)** | **Change** | ***t*(13)** | ***p*** | **Holm *p*** | **d_z** |
|---|---|---|---|---|---|---|---|
| Apply AI (6) | 36.64 (8.88) | 53.00 (7.16) | 16.36 | 5.82 | < .001 | < .001 | 1.56 |
| Understand AI (6) | 33.93 (11.65) | 51.64 (6.28) | 17.71 | 6.02 | < .001 | < .001 | 1.61 |
| Detect AI (3) | 17.57 (6.21) | 24.71 (4.50) | 7.14 | 3.49 | = .004 | = .016 | 0.93 |
| AI Ethics (3) | 20.29 (6.12) | 25.93 (2.87) | 5.64 | 3.33 | = .005 | = .016 | 0.89 |
| Create AI (4) | 6.07 (4.65) | 20.43 (11.82) | 14.36 | 4.64 | < .001 | = .003 | 1.24 |
| AI Problem Solving (3) | 15.14 (6.72) | 25.00 (4.74) | 9.86 | 4.61 | < .001 | = .003 | 1.23 |
| AI Learning (3) | 13.21 (6.86) | 22.64 (5.85) | 9.43 | 6.17 | < .001 | < .001 | 1.65 |
| AI Persuasion Literacy (3) | 25.43 (4.73) | 25.86 (3.72) | 0.43 | 0.32 | = .753 | = .753 | 0.09 |
| AI Emotion Regulation (3) | 23.64 (6.16) | 26.50 (3.41) | 2.86 | 2.46 | = .029 | = .057 | 0.66 |

*Note.* Dimension scores are sums of 0-10 item responses; the number of items appears in parentheses. Change was calculated as Post - Pre. d_z is the paired-samples standardized mean change.

Figure 2

MAILS dimension scores at pretest and posttest

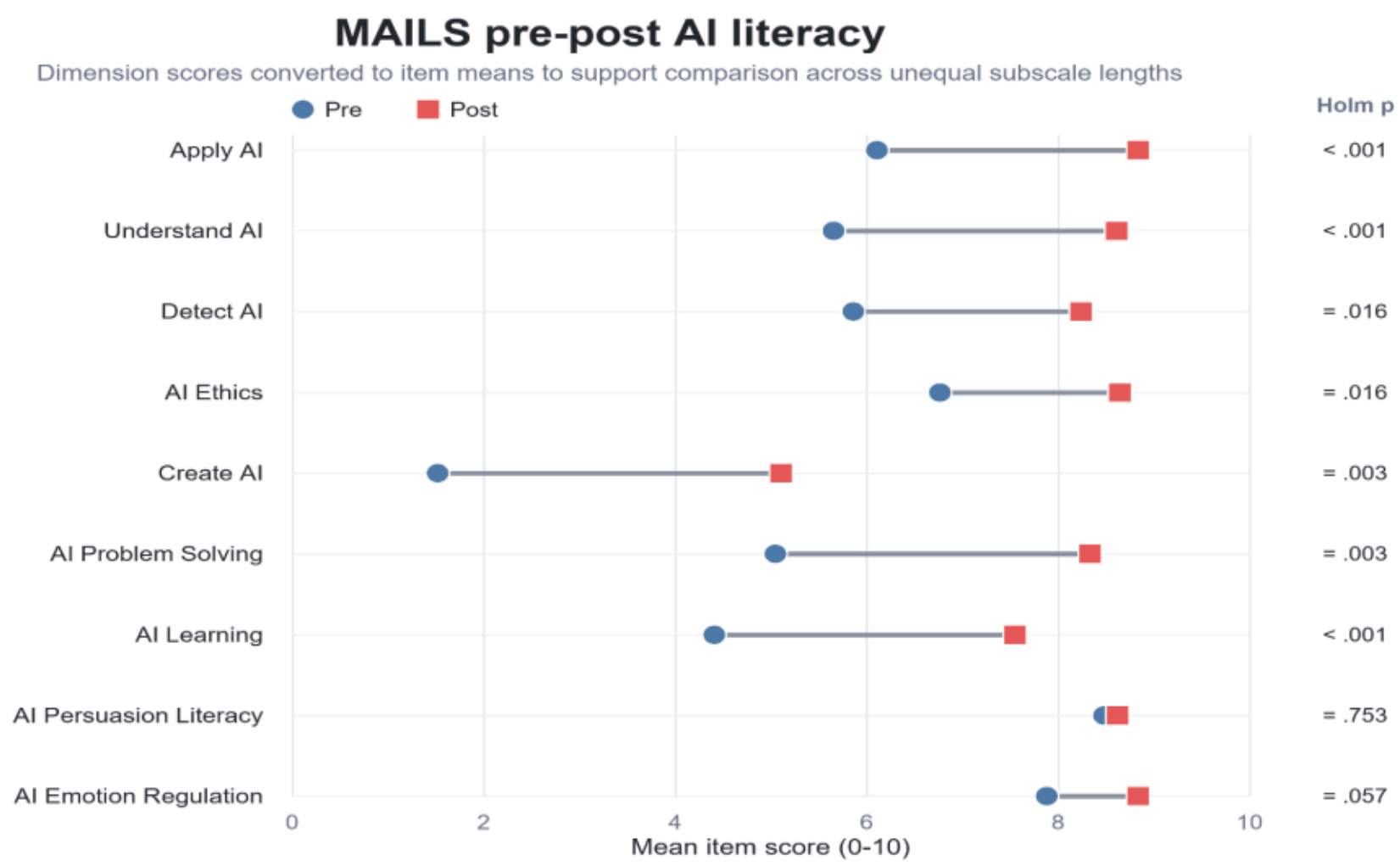

*Note.* Summed dimension scores were divided by the number of items so that all dimensions are displayed on the common 0-10 item-response scale. Holm-adjusted p values are shown at right.

4.1.4 WGCTA (Critical Thinking) Pre-Post Comparison

Wilcoxon signed-rank tests were conducted on WGCTA subscale percentiles given the small sample size and non-normal distribution. Across all three subscales (Draw Conclusions, Evaluate Arguments, Recognize Assumptions), pre-post differences were not statistically significant (all $p > .05$). Mean percentile scores were relatively stable from pre to post test, with no large shifts in central tendency, although standard deviations indicated considerable variability across participants (Table 4).

Table 4

WGCTA Subscale Percentiles+ Figures 3a-c: WGCTA Boxplots, Means, and Paired Differences

WGCTA subscale percentiles and paired Wilcoxon results

| **Subscale** | **Pre M (SD)** | **Post M (SD)** | **Change** | **V** | **p** | **r_rb** |
|---|---|---|---|---|---|---|
| Draw Conclusions | 44.43 (28.95) | 36.71 (29.77) | -7.71 | 31.5 | = .199 | -0.40 |
| Evaluate Arguments | 36.14 (20.80) | 35.86 (31.75) | -0.29 | 46.0 | = .987 | 0.01 |
| Recognize Assumptions | 32.71 (28.50) | 36.50 (29.76) | 3.79 | 65.0 | = .451 | 0.24 |

*Note.* N = 14 matched pairs. Change was calculated as Post - Pre. V is the positive-rank sum, and r_rb is the matched-pairs rank-biserial correlation. Exact permutation p values account for the observed rank ties; conclusions were unchanged after Holm correction.

Figure 3

WGCTA subscale distributions, means, and paired differences

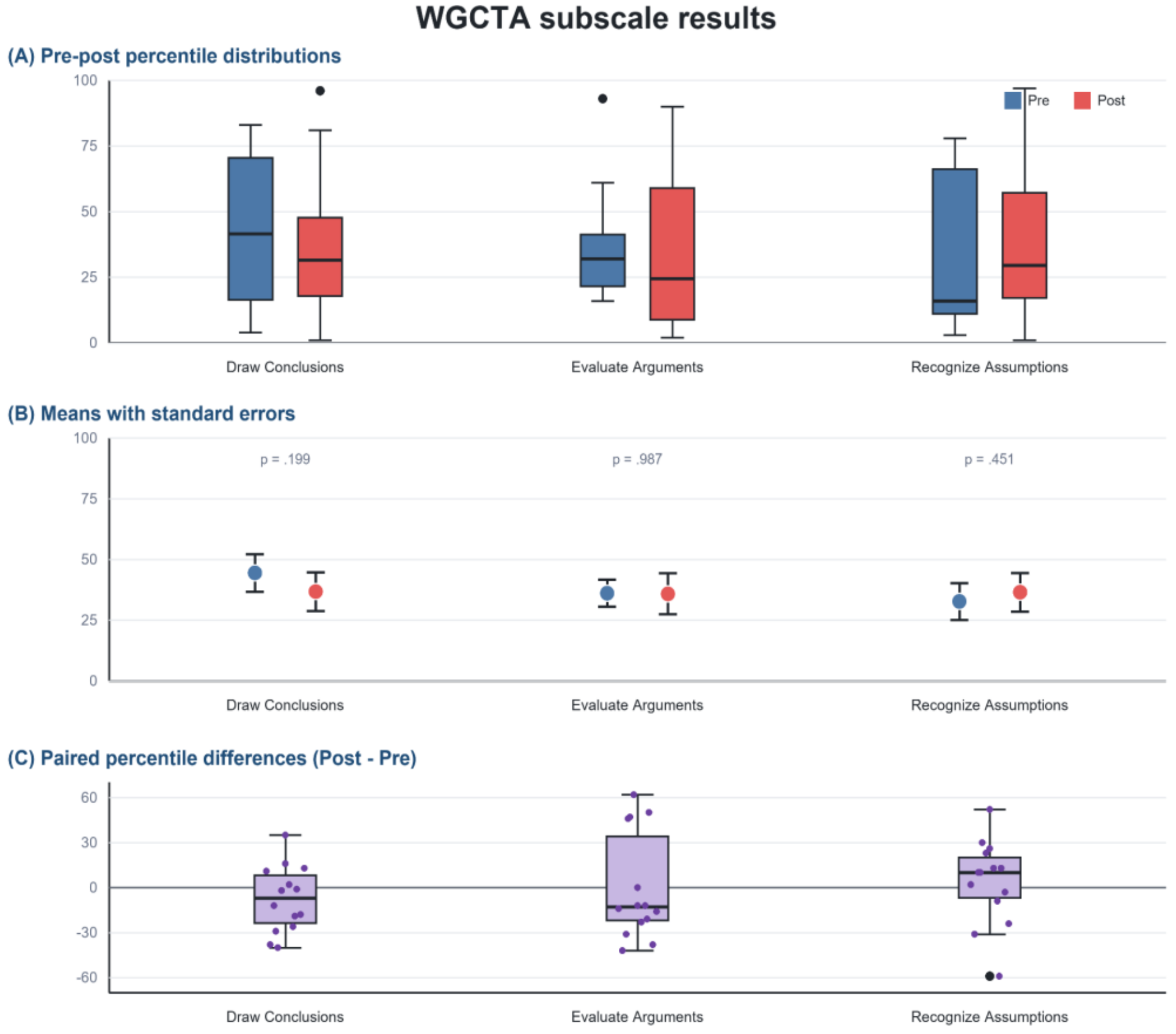

*Note.* Panel A shows pretest and posttest percentile distributions. Panel B shows means with standard errors and unadjusted Wilcoxon p values. Panel C shows paired posttest-minus-pretest differences; the horizontal zero line represents no change.

The pattern of stable WGCTA scores alongside substantial AASCDM and MAILS gains warrants attention. Students self-reported greater statistical and AI competence and showed stronger AI literacy on a validated scale, but they did not show standardized critical thinking gains. We return to this pattern in Section 5.

### 4.2 Patterns in Student-AI Interaction

Topic modeling of the 442 prompts yielded nine distinct thematic clusters (Figure 4). These spanned statistical methods (chi-square association testing, statistical hypothesis testing, applying z-scores, correlation and regression), research design concepts (sampling bias, outlier detection, confidence intervals), and meta-discourse (requesting feedback, research design critique). The clusters separated cleanly, indicating that students raised conceptually distinct topics. Some were concentrated in particular assignments—sampling bias in Assignment 2, hypothesis testing in Assignments 4 and 5—but most contained messages from several assignments, and the requesting-feedback cluster drew from all seven. Core statistical concepts thus recurred across the semester alongside assignment-specific emphases.

Figure 4

*Two-dimensional projection of student prompt embeddings, colored by topic (left) and by assignment (right)*

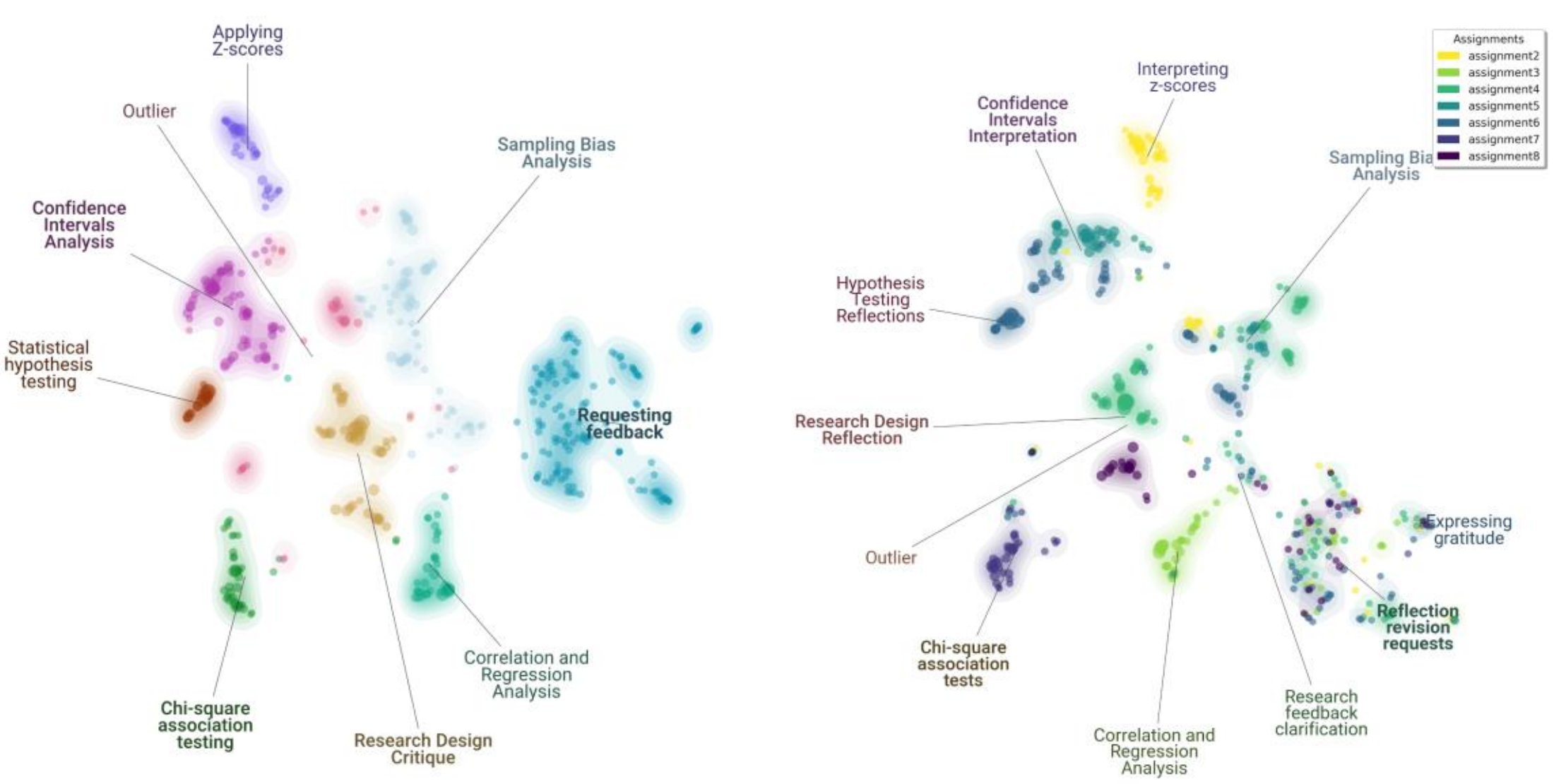

*Note.* Each point is one student message; point size is proportional to message length. Labels in the left panel are the topic representations generated by GPT-4.1.

The coding schema comprised eight categories in two groups. Information Requests/Needs—general improvement request, specific improvement request, reflection analysis, applicability of method, and higher-order thinking—captured inquiries ranging from broad requests for feedback to conceptual questions. Context Description—course description, assignment description, and appreciation—captured messages that supplied background or acknowledgment without requesting substantive help. Against the 50-prompt gold standard, the GPT-5 classifier reached 82% accuracy (41/50, 95% CI [.69, .91]).

Across the full corpus, assignment description was the most common category (36.9%, *n* = 163), followed by general improvement request (22.0%, *n* = 97), reflection analysis (14.3%, *n* = 63), higher-order thinking (11.5%, *n* = 51), appreciation (5.4%, *n* = 24), applicability of method (5.0%, *n* = 22), specific improvement request (3.2%, *n* = 14), and course description (1.8%, *n* = 8). Context description accounted for 44.1% of messages and information requests for 55.9%. These proportions were largely stable across the seven assignments: assignment description and general improvement request remained dominant throughout, and the remaining categories fluctuated within narrow bands with no consistent trend or phase shift (Figure B1).

Students tended to stay in one mode instead of moving between modes. The ten most frequent bigrams account for 65.8% of all turn-to-turn transitions (Table 5), and the four largest are within-category repeats—assignment description to assignment description (21.9%), general improvement request to general improvement request (10.1%), higher-order thinking to higher-order thinking (6.0%), and reflection analysis to reflection analysis (4.6%)—which together make up 42.6% of all transitions and indicate multi-turn elaboration on a single task. The first-order Markov chain shows the same structure as strong diagonal self-loops (assignment description .58, higher-order thinking .47, general improvement request .46; Figure B2). The most frequent cross-category moves ran from supplying context to seeking help (assignment description to general improvement request, 6.8%) and between reflection and conceptual questioning (reflection analysis to higher-order thinking, .22; higher-order thinking to reflection analysis, .21), a reflective–conceptual loop.

Table 5

*Ten most frequent bigram transitions between prompt categories*

| **Transition** | **Count** | **%** |
|---|---|---|
| assignment description → assignment description | 80 | 21.9 |
| general improvement request → general improvement request | 37 | 10.1 |
| assignment description → general improvement request | 25 | 6.8 |
| higher-order thinking → higher-order thinking | 22 | 6.0 |
| reflection analysis → reflection analysis | 17 | 4.6 |
| assignment description → reflection analysis | 14 | 3.8 |
| general improvement request → assignment description | 13 | 3.6 |
| reflection analysis → higher-order thinking | 12 | 3.3 |
| general improvement request → reflection analysis | 11 | 3.0 |
| higher-order thinking → reflection analysis | 10 | 2.7 |

*Note.* Percentages are of all 366 turn-to-turn transitions observed in the corpus. The ten patterns listed account for 65.8% of transitions.

Mapping category flows across the first five conversation positions (Figure 5) reveals three structures. First, categories persist, particularly assignment description, as students progressively supply more detail about their work. Second, a two-phase pattern recurs in which students establish context and then request feedback. Third, transitions from concrete categories (assignment description, general improvement request) to analytical ones (reflection analysis, higher-order thinking) become more frequent at later positions, suggesting that longer conversations afforded some movement toward analytical engagement. This movement was not the norm, however. A substantial share of conversations cycled between describing the assignment and requesting feedback without deepening, which is the interaction-level signature of the validator positioning described in Section 4.3.

Figure 5

*Flow of student prompt categories across the first five conversation positions*

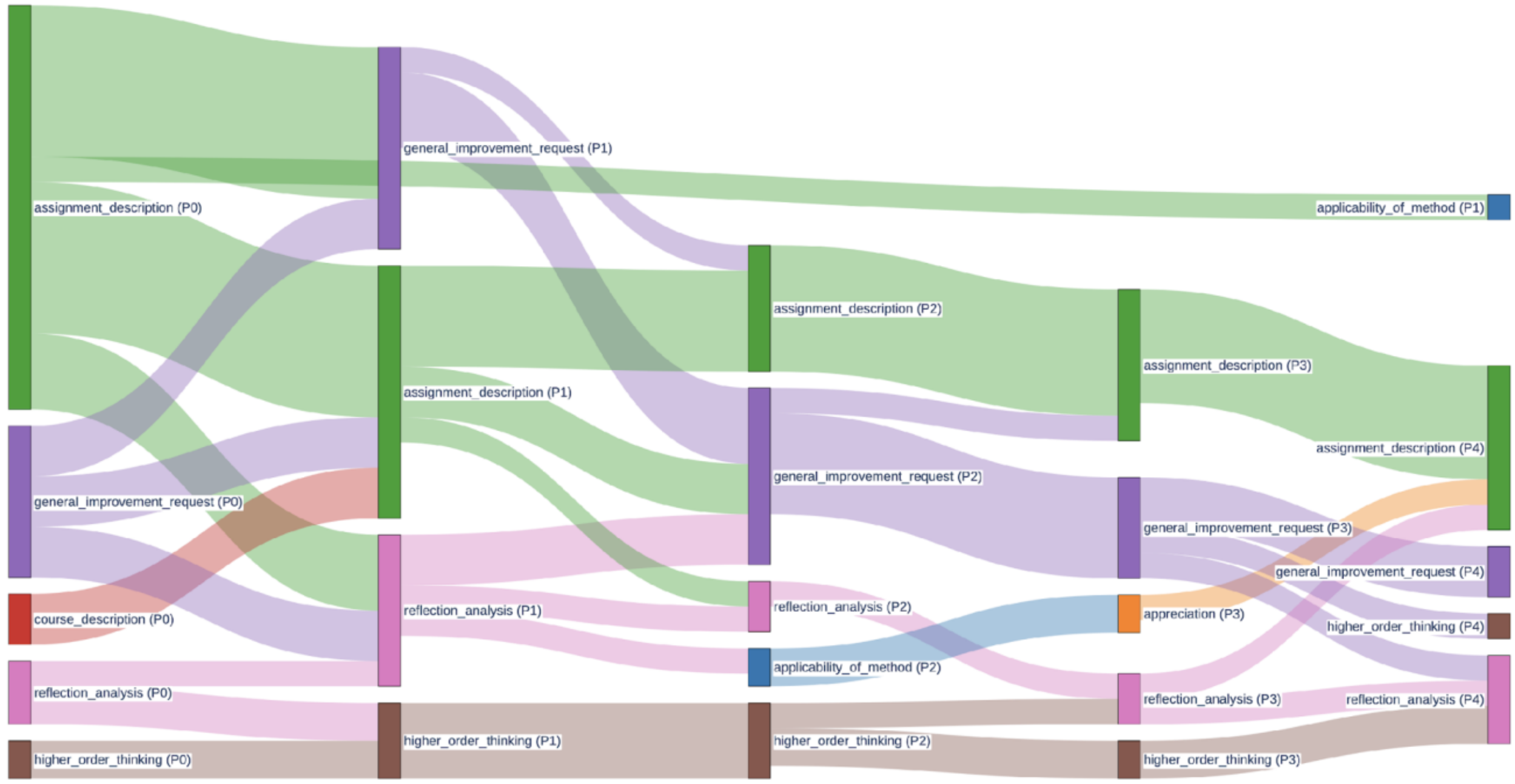

*Note.* Node size and flow width are proportional to frequency; color is consistent for a category across positions (P0–P4). Only patterns occurring at least twice are shown.

## 4.3 Qualitative Reflections from the Design Process

Four interrelated themes emerged from the qualitative analysis: (1) different ways students positioned the LLM, (2) patterns of engagement reflecting deep or conceptual versus shallow or procedural orientations, (3) discursive patterns that occasionally humanized the LLM, and (4) ways the curriculum design itself shaped student engagement. These themes are not separable, but they are not redundant either. For example, the way a student positioned the LLM could shape whether they engaged with it in deep or shallow ways.

We observed three positionings of the LLM: as an answer generator, as an answer validator, or as a co-thinker. Students who positioned it as an answer generator were definitionally displacing their critical thought instead of reinforcing it, with a concomitant negative effect on potential learning. This pattern arose only among a small proportion of participants but signals that future iterations should mitigate it more deliberately. Students who positioned the LLM as an answer validator were not displacing their thinking but also were not visibly building on it. This was the most common positioning and indicates a need to shape future iterations toward more substantive engagement. Finally, students who engaged with the LLM as a co-thinker were actively practicing critical thinking through their interactions, which is the orientation the design is trying to cultivate.

Simultaneously, we observed varying levels of reflective, more conceptual interactions versus shallower, more procedural ones. Eliciting the former is one goal of this curricular intervention, both because of evidence in the learning literature and because critical thinking lives primarily in reflective, more conceptual thought. AI literacy itself can be deep and conceptual or shallow and procedural, and connecting to the first theme above, students who positioned the AI as a co-thinker were positioned to develop deeper AI literacy than those who positioned it only as a generator or validator.

Although we engineered initial prompts intended to give the LLM a warm and empathetic tone, instructors were also clear with students that the LLM is just a machine and cannot truly think; the students must do the thinking. “Smart humans, simple machines” was a common refrain in class. Even so, students sometimes engaged with the LLM in humanized terms, thanking it or sharing personal context such as “I’m tired today.” There is nothing intrinsically wrong with this register, but we note it as something to handle carefully. Discursively constructing the machine in humanized terms does not necessarily mean the student takes it to be sentient, but we want to forestall the lack of critical AI literacy that a presumption of sentience would entail.

Finally, the pilot curriculum’s design shaped each of the patterns above, and revisions to that design are needed across additional design cycles in future semesters and other classes. For example, modifying rubrics to account for broader patterns of LLM interaction could encourage more students to position the LLM as a co-thinker instead of a validator, and incorporating activities that surface the model’s statistical rather than sentient behavior could reinforce that distinction.

### 4.3.1 Interaction Patterns Behind the Positionings

The three positionings identified above corresponded to visibly distinct interaction patterns in the chat logs. The most common pattern was answer-submission for validation: students pasted their drafted responses into the chat and asked the LLM to check them. Common openings included "are all of my answers correct?", "is there anything wrong?", "were my calculations correct", and "are my answers clear enough". Some students pasted their complete draft as a single block; others provided one answer at a time for finer-grained feedback. Both approaches yielded useful information from the LLM, but neither required students to work with what they received. In many submissions the LLM's response was copy-pasted after the student's own answer with no visible integration, consistent with the validator positioning and with reports elsewhere of students treating AI feedback as an artifact to display instead of an input to revise.

A second pattern reflected the generator positioning. Some students pasted the assignment's reflection question directly and asked the LLM to answer it. The prompts were minimal: "why are z-scores important?", "how might z-scores impact my life", "why is a paired t-test perfect?". Here the LLM produced sophisticated responses that a student could paste back into the submission, but the intellectual work of engaging the question had been offloaded to the model. This pattern was less common than validator-style submissions but was clearly present.

The co-thinker positioning appeared in a smaller subset of interactions, typically when students asked substantive follow-up questions or posed original problems instead of assignment prompts. Illustrative student prompts included "if a husband and wife are not on the same page on how satisfied they are, what could explain that", "should AI in therapy be used in a supplementary manor or as a replacement for human therapists", "how could someone misinterpret confidence intervals if they don't know what they are", and "what is the most ethical way to report results collected from purely a self-reported study". Multi-turn sequences within this positioning showed additional signs of engagement: one student asked "chat, should we trust a shorter version of a long, proven test if the scores end up being similar", then followed with "if a test covers 84% of the variability, should it be used to replace it", and later "can outliers affect a sample of 12 more than a sample of 1,000?" These sequences show a student building on the LLM's responses to sharpen the underlying question, which is the pattern the intervention was designed to elicit.

Humanization of the LLM appeared across positionings and was not confined to any single interaction type. Students routinely opened with "Hi there" or "hey chat" and closed with "thank you for your help" or "thank you for your insight". One student wrote, "Thank you! I just put alot of work into that assignment and im really tired but I appreciate the feedback and will probably explore what you said late[r]", a message that combined gratitude, apology, and a personal disclosure of fatigue. This register is not intrinsically problematic, but it does invite closer attention in future iterations: a student who addresses the LLM as they would a supportive teaching assistant may be less inclined to critique its outputs than one who treats it as a statistical model.

One additional pattern warrants attention as a design consideration, not a student behavior. The LLM's own responses frequently concluded with a closed yes-or-no question ("Does that make sense?", "Would you like me to explain further?"). Students often replied to these with a short affirmative and moved to the next question, ending what could have been a productive line of inquiry. The pattern suggests that the LLM's response format was itself shaping the interaction toward validator-style closure. Adjusting the system prompt to elicit open-ended follow-up questions is one revision under consideration for the next iteration. The interaction patterns described in this subsection are consistent with those reported in a companion analysis of the same pilot conducted under a different coding framework (see Discussion below).

### 4.4 Synthesis Across Quantitative and Qualitative Findings

Read together, the quantitative and qualitative findings indicate what one semester of GenAI-assisted instruction can and cannot do. Students made gains in self-assessed statistical learning and on a validated AI literacy scale, including in dimensions that tap conceptual understanding (Understand AI), application (Apply AI), and ethical reasoning (AI Ethics). At the same time, gains on a standardized critical thinking measure did not appear, and the qualitative analysis showed the most common student-LLM relationship was that of validator instead of co-thinker.

The pattern is informative for design. Where the curriculum gave students structured opportunities to use the LLM in well-scoped, course-anchored tasks, gains were strongest. Where the curriculum left student-LLM interaction more open and discretionary, students defaulted to validator-style engagement that did not stretch their reasoning enough to register on a transfer-oriented critical thinking measure. The single MAILS dimension that did not move significantly, AI Persuasion Literacy, asks for the same kind of evaluative reasoning that the WGCTA targets, and the parallel reinforces the inference: the harder skills to develop are those that require sustained, deliberate instructional design instead of exposure alone.

## 5 Discussion

The pilot was designed to ask whether an intentionally Socratic deployment of GenAI inside a research methods and statistics course could support domain learning, AI literacy, and critical thinking together. With N = 14, the answer is partial. Students made gains on every AASCDM dimension and on eight of nine MAILS dimensions. They did not show gains on a standardized measure of critical thinking. The qualitative analysis explains why these two findings appeared together, and points to specific design moves for the next iteration.

The gains in domain learning and AI literacy were strongest in dimensions tied directly to course tasks (interpreting variability, selecting tests, applying AI tools to a defined problem, and so on). The pattern is consistent with prior work suggesting that GenAI can accelerate access to procedural content and applied skill (Lee et al., 2025; Lim et al., 2025) when use is scaffolded by an instructor. For research on adaptive and personalized learning, it suggests that a discourse-partner deployment of GenAI is a useful tool for these outcomes within a single semester.

The flat WGCTA result is the central finding to interpret, and we read it carefully. It is consistent with three explanations that are not mutually exclusive. The first is methodological: transfer-oriented standardized measures of critical thinking are intentionally insensitive to course-specific content, and one semester is a short window for measurable transfer, particularly for reasoning skills that require sustained instruction and deliberate practice to generalize across domains (Bailin, 2002; Halpern, 1998; Niu et al., 2013). The second is behavioral: the most common student-LLM positioning we observed was that of validator, an orientation in which the LLM confirms a student's prior thinking instead of challenging it. The validator pattern is functionally similar to the overreliance described by Zhai and Wibowo (2024) and to the redistribution of critical thinking toward verification observed by Lee et al. (2025). Students confirmed their thinking with the LLM but did not stretch it. The third is curricular: the pilot did not include a dedicated assessment of how students critique the LLM's outputs, which is the part of AI-literate critical thinking that the WGCTA can detect. The next iteration of the framework will add explicit critique-of-AI tasks and matched assessments.

The validator default looks more like a feature of the design than a deficit of the students. Across the qualitative themes, the modal student-LLM relationship was conservative: students confirmed without challenging. This is the predictable pattern of an interaction with a tool that is helpful, polite, and usually correct. Moving students toward the co-thinker positioning we are designing for is not a matter of exhortation. It calls for rubrics that reward visible disagreement with or extension of the LLM's contributions, tasks that surface model limitations and statistical behavior, and classroom routines that normalize naming and resolving moments where the model is wrong or unsure. Each of these is a concrete revision under way for the next DBR iteration.

The MAILS result reinforces the same point. The one dimension that did not move was AI Persuasion Literacy, which asks students to reason about how AI systems may influence belief and behavior. AI Persuasion Literacy is a higher-order, transfer-heavy construct that closely resembles the WGCTA target. The flat result on this MAILS dimension and the flat result on WGCTA are likely the same finding viewed through two instruments. The convergence points to a clear design priority: the next iteration must include direct instruction and authentic practice in evaluating AI outputs for bias, framing, and rhetorical influence.

### 5.1 Design Principles for Constraint-First GenAI Instruction

The pilot generated six design principles for constraint-first GenAI instruction. These are not general prescriptions for all AI-in-education contexts; they are what a semester of teaching this way, analyzing what students actually did, and revising in response taught us.

First, the LLM instruction — the system prompt — has to be tailored to the course and to specific assignments ahead of time. Improvising a general prompt at the start of the semester and asking students to adapt it themselves does not work. The prompt is part of the instructional design and needs to be treated that way.

Second, reflection cannot live in one assignment while the others coast on procedure. In the pilot, the assignments that most consistently produced co-thinker engagement were the ones that built reflection into the task itself. In the next iteration, we integrated reflection components across most of the assignments so that co-thinker engagement had somewhere to land every week, not just when we asked for it explicitly.

Third, the assessment rubric has to reward the thinking, not the answer. When a student submits their AI interaction record alongside their work, and the rubric gives points for the quality of the prompts and follow-ups they authored, the incentive to use the LLM as a validator collapses. Students engage with the LLM more substantively when they know the engagement itself is what is being graded.

Fourth, instructor preparation ahead of the semester is a prerequisite, not a nice-to-have. Constraint-first design asks instructors to redesign assignments, calibrate rubrics, and think through where the LLM belongs in each activity. This work cannot be done during the semester it is being taught. The pilot succeeded partly because the redesign happened before the course began.

Fifth, instructors implementing this design need weekly support to share what they are seeing and to adjust their practice based on student reflections. The pilot was a single instructor iterating on their own, and even then, the mid-semester revisions described in Section 3.2.2 depended on ongoing analysis of what students were producing. A multi-instructor implementation needs a structured space for that same iterative work to happen in near-real-time.

Sixth, student reflections are evidence for design, not just evidence of learning. The most useful signal in the pilot came from what students wrote about how they were engaging with the LLM, and those reflections drove the two mid-semester revisions. Instructors who treat student reflections as diagnostic — as data about whether the design is working — will iterate faster than instructors who treat reflections as an assignment to grade.

These six principles are being tested in a larger implementation currently underway at the authors' institution. That study will report on which principles hold across instructors and disciplinary contexts and which need further refinement.

Several methodological caveats matter. The sample is small (N = 14), drawn from a single course at a single institution, and demographically skewed (predominantly women, predominantly upper-middle income, no first-generation students, no participants with dependents). The participating students were also self-selected into a course that uses GenAI explicitly. We treat all quantitative results as descriptive, not confirmatory and report them to inform the next DBR iteration, not to estimate population effects. The AASCDM, an author-developed self-report instrument, showed acceptable internal consistency at pretest but a ceiling-driven drop at posttest, and the self-report items themselves may inflate perceived gains relative to what a behavioral assessment would show. A companion analysis of the same pilot dataset, conducted under a different coding framework, is under review at another journal; the two analyses report distinct contributions from partially overlapping data, and readers of both should treat their converging findings as views of the same phenomenon through different lenses, not as independent replication. Future iterations will broaden enrollment, span multiple courses, and include comparison sections.

## 6 Conclusion and Implications

This Design-Based Research pilot of a GenAI-assisted critical thinking framework supports three claims. A structured, Socratic deployment of ChatGPT inside a research methods and statistics course produced gains in domain learning and most dimensions of AI literacy within a single semester. The same deployment did not produce gains on a standardized measure of critical thinking, and the qualitative evidence indicates that one reason is that most students engaged with the LLM as a validator instead of a co-thinker. The pilot also points to the design moves required to address that gap: rubrics that reward visible critique of AI outputs, tasks that surface model limitations, and classroom routines that normalize disagreement with the LLM.

For research on adaptive and personalized learning, the pilot offers a case in which GenAI's adaptivity was used as a Socratic thinking partner in place of content delivery or recommendation. The framing has both promise and demands. It is promising because the underlying interaction is closer to the kind of dialogic learning that decades of educational research support. It is demanding because the work of getting students to a co-thinker orientation falls on instructional design, not the model itself. Personalization in this framing is not algorithmic content matching; it is the model adapting its register and depth to a particular student in a particular conversation, with the instructor designing the conditions under which that adaptation is productive.

For practice, three implications follow. Instructors integrating GenAI into critical thinking courses should plan for explicit instruction in AI persuasion literacy and direct critique of AI outputs instead of assuming these will emerge from exposure. They should design rubrics that make co-thinker engagement visible and gradeable. And they should pair self-report measures of AI literacy with observation of actual AI interactions, because the two are related but not interchangeable.

For policy and institutional decision-making, the pilot suggests that GenAI deployments in higher education will succeed or fail not on the basis of model choice but on the basis of pedagogical scaffolding. Institutions investing in AI access without parallel investment in faculty development and curriculum design are likely to see validator-style use predominate, with attendant displacement of the reflective learning the technology was supposed to support.

This Design-Based Research pilot used a constraint-first framing to develop a GenAI-mediated learning design intended to preserve and stimulate higher-order thinking, including critical thinking. An initial iteration with 14 undergraduates was analyzed for student activity, work products, and implementation evidence; the resulting design principles are informing a larger, multi-course college-level implementation currently underway at the authors' institution. Future work will report on the next iterations of the framework, including new tasks that target AI persuasion literacy directly, expanded sample sizes across disciplinary contexts, and objective measures of higher-order thinking to complement the standardized assessments used here.

## Data Availability Statement

The de-identified interaction log corpus and analysis code that support the findings of this study are available from the corresponding author on reasonable request, subject to the terms of the study's Institutional Review Board approval. Raw interaction logs cannot be shared publicly because participant consent was obtained for research use only.

## Declaration of Generative AI and AI-Assisted Technologies

The authors used GPT-4.1 and GPT-5 (OpenAI) as research instruments for the topic modeling representation labels and prompt classification described in Section 3.4; which is part of the research design and is described in Methods. During the preparation of this work, the authors used ChatGPT (OpenAI) and Claude (Anthropic) to assist with grammar checking, readability, reference formatting, and cross-checking in-text citations against the reference list.

After using these tools, the authors reviewed and edited all content and took full responsibility for the content of the published article.